\documentclass[aps,prd,reprint,superscriptaddress]{revtex4-2}

\usepackage[T1]{fontenc}
\usepackage[utf8]{inputenc}
\usepackage{amsmath}
\usepackage{amssymb}
\usepackage{bm}
\usepackage{graphicx}
\usepackage{slashed}
\usepackage[colorlinks=true,linkcolor=blue,citecolor=blue,urlcolor=blue]{hyperref}

\newcommand{\tr}{\operatorname{tr}}

\begin{document}

\title{Pathways to Real Composite Operators from Non-Hermitian Fermions}

\author{V.~E.~R.~Lemes}
\email{verlemes@gmail.com}
\affiliation{Instituto de F\'isica, Universidade do Estado do Rio de Janeiro,
Rua S\~ao Francisco Xavier 524, Maracan\~a, Rio de Janeiro - RJ, 20550-013, Brazil}

\author{D.~G.~Tedesco}
\email{daniel.te@uninter.com}
\affiliation{ESEHL e PPGENT, Centro Universit\'ario Internacional UNINTER,
Av. Sete de Setembro 4228, Batel, Curitiba - PR, 80250-085, Brazil}

\date{\today}

\begin{abstract}
A field theory in $3+1$ dimensions contains two fermions, two Abelian gauge fields, and one complex scalar, with dynamics fixed by a BRST symmetry. For a specific parameter configuration, the fermion mass matrix becomes non-Hermitian, and the propagators exhibit complex conjugate poles. We evaluate the one-loop fermion contribution to the scalar two-point function. Once the factor $i$ of the $e^{iS}$ normalization is removed, the contribution is real for real external momentum below the threshold $(m+m^{*})^{2}$ of the two constituents, which follows from the pairing of complex conjugate terms in the loop integral. The thresholds of the equal-mass pairs sit at complex $p^{2}$ and generate no cut on the real axis, while the mixed pair, whose constituent masses add up to the real number $m+m^{*}$, opens a cut. Across this cut the upper and lower boundary values are related by complex conjugation, so that the associated spectral density is real. Its positivity is not established here. The BRST construction of the action provides the starting point for an algebraic renormalization analysis.
\end{abstract}

\maketitle

\section{Introduction}

Non-Hermitian Hamiltonians describe consistent quantum systems when an antilinear symmetry organizes the spectrum. Many $\mathcal{PT}$-symmetric Hamiltonians have a real spectrum \cite{Bender1998,Bender2007}. The mechanism behind this is pseudo-Hermiticity. A Hamiltonian that obeys $\eta H\eta^{-1}=H^{\dagger}$, with $\eta$ Hermitian and invertible, has a real spectrum and evolves unitarily under the inner product set by $\eta$ \cite{Mostafazadeh2002,Mostafazadeh2010}. Antilinear symmetry controls the reality of the spectrum and the positivity of that inner product, and it yields unitary evolution when $H$ is not self-adjoint in the Dirac sense \cite{Mannheim2013}.

Complex eigenvalues are admitted in this setting. Complex conjugate pairs in the spectrum produce propagators with complex conjugate poles. Lee and Wick found such poles in an electrodynamics with an indefinite-metric sector that cancels ultraviolet divergences, where the extra excitations acquire complex masses \cite{LeeWick1969}. The same poles arise in the infrared of confining gauge theories. The Gribov-Zwanziger gluon propagator $k^{2}/(k^{4}+\gamma^{4})$ factorizes into modes of complex mass $\pm i\gamma^{2}$, the $i$-particles \cite{iparticles,Sorella2011}, and the refined Gribov-Zwanziger action sharpens this picture in agreement with the lattice \cite{RGZ}. The same complex-pole structure extends to the matter sector through a soft breaking of the BRST symmetry, whose multiplicative renormalizability has been established to all orders of perturbation theory \cite{Baulieu2010,Capri2011, Capri:2011ki}. Composite operators built from such complex-pole constituents have correlators with a real spectral representation \cite{iparticles,Dudal:2010wn}, which also yields glueball mass estimates in agreement with the lattice \cite{DudalGlueball2011}. The non-Hermiticity sits in the elementary propagators, while suitable composite correlators can admit real spectral representations.

Non-Hermitian fermion sectors follow related constructions. A Dirac fermion with a parity-odd anti-Hermitian mass and a non-Hermitian Yukawa coupling produces light neutrino masses and an extra source of $CP$ violation \cite{ABM2015}. Spontaneous symmetry breaking and the Goldstone theorem have been worked out for non-Hermitian scalar and gauge theories \cite{AEMS2018}. Non-Hermitian dynamics also appears in condensed matter, from $\mathcal{PT}$-symmetric dimers to exciton-polariton systems \cite{Ge2024,Xu2024}.

The model studied here places a non-Hermitian fermion mass matrix in a Yukawa and Higgs sector governed by a BRST symmetry. The fields are two Abelian gauge fields $a_{\mu}$ and $\hat{a}_{\mu}$, two fermions $\psi$ and $\chi$, and a complex scalar $\phi$. A quartic fermion interaction, non-renormalizable by power counting, becomes local once an auxiliary scalar enters through a Hubbard-Stratonovich rewriting \cite{Amit:1978dk,Zinn-Justin:1996khx,Kleinert:2011rb}, and BRST invariance fixes the covariant derivatives and the admissible terms \cite{Piguet:1995er}. The scalar potential breaks the symmetry spontaneously, $\phi$ acquires a vacuum expectation value $v$, and the Yukawa terms turn into mixed fermion masses. For $a=-b$ with $a,b\in\mathbb{R}$ the mass matrix in the $(\psi,\chi)$ basis is non-Hermitian, with complex conjugate eigenvalues $m_{\pm}=N\pm iav$, so the fermion propagators carry complex poles.

The one-loop fermion contribution to the scalar two-point function $\langle\phi^{\dagger}(p)\phi(-p)\rangle$ is real for real external momentum below $(m+m^{*})^{2}=4N^{2}$, in the normalization set by $e^{iS}$, and the reality follows from the pairing of complex conjugate terms in the loop integral. Above that value a cut opens, since the masses of the mixed pair add up to a real number even though each of them is complex; the upper and lower boundary values are then exchanged by complex conjugation and define a real spectral density, while the equal-mass thresholds, sitting at $4m^{2}$ and $4(m^{*})^{2}$, contribute no singularity on the real axis. The model carries the $i$-particle pattern into a Yukawa and Higgs setting, with complex poles in the constituent fermions and a real scalar loop coefficient below threshold together with a real spectral density on the cut. The action is BRST invariant and power-counting renormalizable, which provides the starting point for a full algebraic renormalization analysis. The metric that makes the evolution unitary on a suitable subspace is left for later work, and the present one-loop reality result does not by itself establish such a metric.

The remainder of the article develops these statements in sequence. The construction of the model occupies the section that follows, where an auxiliary scalar renders the quartic fermion interaction local, the BRST transformations determine the covariant derivatives together with the admissible action, and spontaneous symmetry breaking converts the Yukawa terms into a fermion mass matrix whose Green's functions are obtained both for the Hermitian alternative $a=b$ and for its non-Hermitian counterpart $a=-b$. The non-Hermitian case then receives a dedicated treatment in the $\Psi_{\pm}$ basis, where the propagators become diagonal and describe Dirac fermions of complex conjugate mass, and where the composite operator $\overline{\Psi}_{+}\Psi_{-}$ combines its constituents into a real spectral representation. From this structure follows the central computation, the one-loop fermion contribution to the two-point function of the scalar field, defined in Euclidean space, carried out by Feynman parametrization and dimensional regularization, and shown to remain real for real external momentum below the threshold of the mixed pair. The analytic structure that underlies this definition, the continuation to real external momentum, and the location of the thresholds are treated in Appendix~\ref{app:euclidean}. The gauge sector completes the analysis, since the scalar expectation value generates the gauge-field masses, the BRST construction supplies the gauge fixing, and the cohomology of the broken vacuum organizes the gauge and ghost propagators that close it. The conclusions gather the results.

\section{Interacting fermions}

A theory of two fermions with a quartic self-interaction has the action
\begin{multline}\label{eq:fourfermion}
\int d^{4}x\,\bigg[\, i\overline{\psi}\gamma^{\mu}\partial_{\mu}\psi + i\overline{\chi}\gamma^{\mu}\partial_{\mu}\chi + N(\overline{\psi}\psi+\overline{\chi}\chi) \\
+ \lambda_{1}\overline{\psi}\chi\overline{\chi}\psi + \lambda_{2}\overline{\chi}\psi\overline{\psi}\chi \,\bigg].
\end{multline}
Power counting makes this action non-renormalizable. The couplings $\lambda_{1}$ and $\lambda_{2}$ have negative mass dimension, and the expansion produces terms of ever higher order in the fermion fields.

An auxiliary scalar localizes the quartic interaction, as in the Hubbard-Stratonovich transformation \cite{Amit:1978dk,Zinn-Justin:1996khx,Kleinert:2011rb}. A symmetry that extends to the quantum level constrains the field content and the admissible terms and secures perturbative stability. The fields carry the BRST transformations \cite{Piguet:1995er}
\begin{equation}\label{eq:brst}
\begin{aligned}
s a_\mu &= -\partial_\mu c, & s \hat{a}_\mu &= \partial_\mu \hat{c}, \\
s c &= 0, & s \hat{c} &= 0, \\
s \psi &= -ie\,c\,\psi, & s \overline{\psi} &= ie\,\overline{\psi}\,c, \\
s \chi &= -ie\,\hat{c}\,\chi, & s \overline{\chi} &= ie\,\overline{\chi}\,\hat{c}, \\
s \phi &= ie\,(\hat{c} - c)\,\phi, & s \phi^\dagger &= -ie\,(\hat{c} - c)\,\phi^\dagger.
\end{aligned}
\end{equation}
The set acts on the fermions, the scalar, the two Faddeev-Popov ghosts, and the two $U(1)$ gauge fields. The covariant derivatives fix the signs of the gauge transformations. The operator $s$ is nilpotent, $s^{2}=0$, and acts as a graded derivation,
\begin{equation}\label{eq:graded}
s(\Phi_1 \Phi_2) = (s\Phi_1) \Phi_2 + (-1)^{|\Phi_1|} \Phi_1 (s\Phi_2),
\end{equation}
the grading $|\Phi|$ being the ghost number, so that $|\Phi|=1$ for the ghosts $c$ and $\hat{c}$ and $|\Phi|=0$ for every other field, the matter fields included. With this grading $s$ commutes past the fermion fields, and the relative signs in \eqref{eq:brst} are the ones that leave $N\overline{\psi}\psi$, $N\overline{\chi}\chi$ and the Yukawa terms invariant.

\subsection{Covariant derivatives}

Homogeneity of the covariant derivatives under BRST,
\begin{equation}\label{eq:covcond}
s(D_\mu \Phi) = ie(\hat{c} - c)(D_\mu \Phi) \quad \text{for } \Phi = \phi, \phi^\dagger,
\end{equation}
fixes the gauge couplings. The scalar covariant derivative that meets this condition reads
\begin{equation}\label{eq:Dphi}
\begin{aligned}
D_\mu \phi &= \big( \partial_\mu - i e (a_\mu + \hat{a}_\mu) \big) \phi, \\
s(D_\mu \phi) &= ie(\hat{c}-c) D_\mu \phi.
\end{aligned}
\end{equation}
The conjugate field carries the Hermitian conjugate,
\begin{equation}\label{eq:Dphidag}
\begin{aligned}
D_\mu \phi^\dagger &= \big( \partial_\mu + i e (a_\mu + \hat{a}_\mu) \big) \phi^\dagger, \\
s(D_\mu \phi^\dagger) &= -ie(\hat{c} - c) D_\mu \phi^\dagger,
\end{aligned}
\end{equation}
consistent with $s\phi^\dagger = -ie(\hat{c}-c)\phi^\dagger$. The scalar and its conjugate couple to the sum $a_{\mu}+\hat{a}_{\mu}$, so each carries the same charge under the two $U(1)$ groups,
\begin{equation}\label{eq:charges}
q_a = -e \ \text{(under $a_\mu$)}, \qquad q_{\hat{a}} = -e \ \text{(under $\hat{a}_\mu$)}.
\end{equation}
The fermion covariant derivatives carry opposite signs, set by the relative charges,
\begin{equation}\label{eq:fermcov}
\begin{aligned}
\nabla_\mu \psi &= \big( \partial_\mu - i e a_\mu \big) \psi, \\
\hat{\nabla}_\mu \chi &= \big( \partial_\mu + i e \hat{a}_\mu \big) \chi,
\end{aligned}
\end{equation}
and one checks
\begin{equation}\label{eq:fermcovBRST}
s(\nabla_\mu \psi) = -ie\,c\, (\nabla_\mu \psi), \qquad
s(\hat{\nabla}_\mu \chi) = -ie\,\hat{c}\, (\hat{\nabla}_\mu \chi).
\end{equation}
In condensed-matter terms the spectrum holds a fermion of charge $-e$ and a second fermion of opposite charge, the analogue of a hole. BRST invariance then allows the action
\begin{widetext}
\begin{multline}
S = \int d^4x\, \Big[\, -\tfrac{1}{4} f_{\mu\nu}(a)f^{\mu\nu}(a) -\tfrac{1}{4} f_{\mu\nu}(\hat{a})f^{\mu\nu}(\hat{a})
+ i\overline{\psi}\gamma^{\mu}\nabla_{\mu}\psi + i\overline{\chi}\gamma^{\mu}\hat{\nabla}_{\mu}\chi \\ + N\overline{\psi}\psi + N\overline{\chi}\chi + a\,\overline{\psi}\phi\,\chi + b\,\overline{\chi}\phi^{\dagger}\psi+(D_\mu \phi)^\dagger (D^\mu \phi) - V(\phi^\dagger \phi) \,\Big],\label{eq:action}
\end{multline}
\end{widetext}
with
\begin{equation}\label{eq:field-strengths}
f_{\mu\nu}(a) = \partial_\mu a_\nu - \partial_\nu a_\mu, \qquad
f_{\mu\nu}(\hat{a}) = \partial_\mu \hat{a}_\nu - \partial_\nu \hat{a}_\mu,
\end{equation}
and
\begin{equation}\label{eq:potential}
V(\phi^\dagger \phi) = -m^2 \phi^\dagger \phi + \frac{\lambda}{2} (\phi^\dagger \phi)^2,
\qquad m^2 > 0, \ \lambda > 0,
\end{equation}
where $m^2>0$ and $\lambda>0$ keep the potential bounded from below. The minimum sits at
\begin{equation}\label{eq:vev}
\langle \phi^\dagger \phi \rangle = \frac{m^2}{\lambda} \equiv v^2, \qquad v = \sqrt{m^2/\lambda}.
\end{equation}

Spontaneous symmetry breaking gives the scalar the expectation values
\begin{equation}\label{eq:ssb}
\langle \phi \rangle = v, \qquad \langle \phi^\dagger \rangle = v.
\end{equation}
The Yukawa sector then produces the mixed mass terms
\begin{equation}\label{eq:mixedmass}
a\,\overline{\psi}\langle \phi \rangle\chi = a v \,\overline{\psi}\chi, \qquad
b\,\overline{\chi}\langle \phi^\dagger \rangle\psi = b v \,\overline{\chi}\psi.
\end{equation}

\subsection{Fermionic bilinear in the $(\psi,\chi)$ basis}

The bilinear terms in the $(\psi,\chi)$ basis form the quadratic fermionic action in momentum space,
\begin{equation*}
S_{\text{fermi}}^{(2)}= \int\!\!
\frac{d^4k}{(2\pi)^4} \begin{pmatrix} \overline{\psi}(k) & \overline{\chi}(k) \end{pmatrix}
\begin{pmatrix} \slashed{k} - N & -a v \\ -b v & \slashed{k} - N \end{pmatrix}
\begin{pmatrix} \psi(k) \\ \chi(k) \end{pmatrix},
\end{equation*}
with $\slashed{k} = \gamma^\mu k_\mu$ and the factor $i$ absorbed into the propagator. The operator to invert reads
\begin{equation}\label{eq:Mfermi}
M_{\text{fermi}}(k) = \begin{pmatrix} \slashed{k} - N & -a v \\ -b v & \slashed{k} - N \end{pmatrix},
\end{equation}
and the Green's functions come from its inverse in Dirac and flavor space. The low-momentum limit $\slashed{k}\to 0$ gives the mass matrix
\begin{equation}\label{eq:Mmass}
M_{\text{mass}} = \begin{pmatrix} N & a v \\ b v & N \end{pmatrix},
\end{equation}
whose eigenvalues are the masses after diagonalization.

\subsection{Fermion Green's functions}

The Yukawa terms produce the mass matrix
\begin{equation}\label{eq:massmatrix}
M = \begin{pmatrix} N & a v \\ b v & N \end{pmatrix},
\end{equation}
and diagonalization gives the fermion Green's functions. A Hermitian theory has $b=a^{*}$ and two real masses. Two real-coupling cases follow.

\subsubsection{Case 1: $a = b$ (real, Hermitian)}

The mass matrix is Hermitian, with real eigenvalues
\begin{equation}\label{eq:case1mass}
m_1 = N + a v, \qquad m_2 = N - a v,
\end{equation}
and the Green's functions read
\begin{equation}\label{eq:G1}
\begin{aligned}
G(\overline{\psi}\psi) &= \frac{i}{2} \left[ \frac{\slashed{k} + m_1}{k^2 - m_1^2} + \frac{\slashed{k} + m_2}{k^2 - m_2^2} \right], \\
G(\overline{\chi}\chi) &= \frac{i}{2} \left[ \frac{\slashed{k} + m_1}{k^2 - m_1^2} + \frac{\slashed{k} + m_2}{k^2 - m_2^2} \right], \\
G(\overline{\chi}\psi) &= \frac{i}{2} \left[ \frac{\slashed{k} + m_1}{k^2 - m_1^2} - \frac{\slashed{k} + m_2}{k^2 - m_2^2} \right], \\
G(\overline{\psi}\chi) &= \frac{i}{2} \left[ \frac{\slashed{k} + m_1}{k^2 - m_1^2} - \frac{\slashed{k} + m_2}{k^2 - m_2^2} \right].
\end{aligned}
\end{equation}

\subsubsection{Case 2: $a = -b$ (real, non-Hermitian)}

The mass matrix is non-Hermitian and its eigenvalues are complex,
\begin{equation}\label{eq:case2mass}
m = N + i a v, \qquad m^{\ast} = N - i a v.
\end{equation}
The Green's functions keep the Case 1 form with complex masses,
\begin{equation}\label{eq:G2}
\begin{aligned}
G(\overline{\psi}\psi) &= \frac{i}{2} \left[ \frac{\slashed{k} + m}{k^2 - m^2} + \frac{\slashed{k} + m^{\ast}}{k^2 - (m^{\ast})^2} \right], \\
G(\overline{\chi}\chi) &= \frac{i}{2} \left[ \frac{\slashed{k} + m}{k^2 - m^2} + \frac{\slashed{k} + m^{\ast}}{k^2 - (m^{\ast})^2} \right], \\
G(\overline{\chi}\psi) &= \frac{i}{2} \left[ \frac{\slashed{k} + m}{k^2 - m^2} - \frac{\slashed{k} + m^{\ast}}{k^2 - (m^{\ast})^2} \right], \\
G(\overline{\psi}\chi) &= \frac{i}{2} \left[ \frac{\slashed{k} + m}{k^2 - m^2} - \frac{\slashed{k} + m^{\ast}}{k^2 - (m^{\ast})^2} \right].
\end{aligned}
\end{equation}
The second case connects to non-Hermitian systems. The Green's functions carry complex poles and do not describe physical particles by themselves, yet a product of fermions with complex poles has a real K\"all\'en-Lehmann spectral representation, and the scalars $\phi$ and $\phi^{\dagger}$ act as the corresponding condensate.

\section{Study of the $a=-b$ case}

For real $a,b$ with $a=-b$ the mass matrix in the $(\psi,\chi)$ basis reads
\begin{equation}\label{eq:Mab}
M = \begin{pmatrix} N & a v \\ -a v & N \end{pmatrix},
\end{equation}
non-Hermitian, with complex conjugate eigenvalues
\begin{equation}\label{eq:eigab}
m = N + i a v, \qquad m^{\ast} = N - i a v.
\end{equation}
Throughout we work in the regime of perturbative $av$,
\begin{equation}\label{eq:regime}
N^{2} > a^{2}v^{2},
\end{equation}
which keeps $m^{2}$ inside the right half plane and, as shown in
Appendix~\ref{app:euclidean}, keeps the Euclidean denominators of the
one-loop integral away from zero.

The combinations $\Psi_{+}$ and $\Psi_{-}$,
\begin{equation}\label{eq:Psidef}
\Psi_+ = \frac{\psi + i \chi}{\sqrt{2}}, \qquad
\Psi_- = \frac{\psi - i \chi}{\sqrt{2}},
\end{equation}
invert to
\begin{equation}\label{eq:Psiinv}
\psi = \frac{\Psi_+ + \Psi_-}{\sqrt{2}}, \qquad
\chi = \frac{\Psi_+ - \Psi_-}{i\sqrt{2}}.
\end{equation}

\subsection{$\mathbb{Z}_2$ symmetry and the pairing of complex poles}

The condition $a=-b$ endows the action with a discrete $\mathbb{Z}_2$ symmetry,
which acts on the fields as
\[
\psi \leftrightarrow \chi,\quad
\chi \rightarrow -\psi,\quad
a_\mu \rightarrow -\hat{a}_\mu,\quad
\phi \leftrightarrow \phi^\dagger,
\]
and similarly for the conjugates.
Applying the transformation twice returns the identity on all physical observables,
since fermions acquire an unobservable minus sign.
This symmetry is exact only when $a=-b$, as follows from the Yukawa sector.

In the $\Psi_\pm$ basis defined in \eqref{eq:Psidef}, the $\mathbb{Z}_2$ symmetry
acts as $\Psi_\pm \to \pm i \Psi_\mp$ up to a phase, and it forces the fermion
propagators to appear in complex conjugate pairs,
$G(\Psi_+\overline{\Psi}_+) \sim (i)/(\slashed{k}-m)$ and
$G(\Psi_-\overline{\Psi}_-) \sim (i)/(\slashed{k}-m^*)$, with $m=N+iav$.
Consequently, the one-loop contribution to $\langle \phi^\dagger(p)\phi(-p)\rangle$
becomes real for real external momentum below the threshold of the mixed
pair, as shown below.
The $\mathbb{Z}_2$ symmetry thus provides the algebraic structure that protects
the reality of composite correlators despite the non-Hermitian mass matrix.

In the $\Psi_{\pm}$ basis the Green's functions are diagonal and describe Dirac fermions of complex conjugate masses,
\begin{equation}\label{eq:Gdiag}
\begin{aligned}
G(\Psi_+ \overline{\Psi}_+) &= \frac{i}{\slashed{k} - m}, \\
G(\Psi_- \overline{\Psi}_-) &= \frac{i}{\slashed{k} - m^{\ast}},
\end{aligned}
\end{equation}
with vanishing mixed components,
\begin{equation}\label{eq:Gmixed}
G(\Psi_+ \overline{\Psi}_-) = G(\Psi_- \overline{\Psi}_+) = 0.
\end{equation}
No $i\epsilon$ is displayed in \eqref{eq:Gdiag}, since the imaginary part of
$m^{2}$ is finite and already fixes the position of the singularities in the
complex energy plane; the analytic structure that results is the subject of
Appendix~\ref{app:euclidean}. Their standard form reads
\begin{equation}\label{eq:Gstd}
\begin{aligned}
G(\Psi_+ \overline{\Psi}_+) &= i\, \frac{\slashed{k} + m}{k^2 - m^2}, \\
G(\Psi_- \overline{\Psi}_-) &= i\, \frac{\slashed{k} + m^{\ast}}{k^2 - (m^{\ast})^2}.
\end{aligned}
\end{equation}
The inverse transformation gives
\begin{equation}\label{eq:Goriginal}
\begin{aligned}
G(\psi \overline{\psi}) &= \frac{1}{2}\big( G(\Psi_+ \overline{\Psi}_+) + G(\Psi_- \overline{\Psi}_-) \big), \\
G(\chi \overline{\chi}) &= \frac{1}{2}\big( G(\Psi_+ \overline{\Psi}_+) + G(\Psi_- \overline{\Psi}_-) \big), \\
G(\psi \overline{\chi}) &= \frac{i}{2}\big( G(\Psi_+ \overline{\Psi}_+) - G(\Psi_- \overline{\Psi}_-) \big), \\
G(\chi \overline{\psi}) &= -\frac{i}{2}\big( G(\Psi_+ \overline{\Psi}_+) - G(\Psi_- \overline{\Psi}_-) \big).
\end{aligned}
\end{equation}

\subsection{Green's function of the composite operator $\overline{\Psi}_+ \Psi_-$}

The product of Green's functions met in loops, as in a vacuum polarization or in the two-point function of $\overline{\Psi}_+ \Psi_-$, reads
\begin{equation}\label{eq:product}
\begin{aligned}
G(\Psi_+ \overline{\Psi}_+)\, G(\Psi_- \overline{\Psi}_-) &= \frac{i^2}{(\slashed{k} - m)(\slashed{k} - m^{\ast})} \\
&= - \frac{(\slashed{k} + m)(\slashed{k} + m^{\ast})}{(k^2 - m^2)(k^2 - (m^{\ast})^2)}.
\end{aligned}
\end{equation}
The Dirac trace gives
\begin{equation*}
\tr\big( G(\Psi_+ \overline{\Psi}_+)\, G(\Psi_- \overline{\Psi}_-) \big) = -4\, \frac{k^2 + |m|^2}{(k^2 - m^2)(k^2 - (m^{\ast})^2)},
\end{equation*}
with $m m^{\ast} = |m|^2 = N^2 + a^2 v^2$. The integral of these products of complex conjugate Green's functions has a real K\"all\'en-Lehmann spectral representation \cite{Dudal:2010wn} in Euclidean space. The dependence on real external momentum follows from the Euclidean expression by analytic continuation, in the sense made precise in Appendix~\ref{app:euclidean}.

\section{Fermion contribution to $\langle\phi^{\dagger}(p)\phi(-p)\rangle$}

The one-loop fermion contribution to the scalar two-point function $\langle\phi^{\dagger}(p)\phi(-p)\rangle$ shows how a scalar two-point kernel can remain real below threshold when the internal Green's functions carry complex conjugate poles. The Yukawa interactions in \eqref{eq:action} fix the one-loop integral.

\textit{Definition of the loop integral.} We state at the outset how the
integral is defined, and we formulate it in Euclidean space from the
start. The complex conjugate masses place the singularities of the two
propagators in opposite quadrants of the complex energy plane, and no
single rotation of the Minkowski contour reaches the Euclidean region in
all the terms that occur, as Appendix~\ref{app:nowick} shows in detail.
Following the treatment of composite operators built from constituents
with complex conjugate poles \cite{iparticles,Dudal:2010wn}, we define the
loop integral in Euclidean space, where the denominators never vanish on
the domain of integration, and we obtain the dependence on real external
momentum by analytic continuation in $p^{2}$. The equal-mass thresholds
are displaced into the complex $p^{2}$ plane, whereas the mixed pair has
the real threshold $p^{2}=(m+m^{*})^{2}=4N^{2}$.
Appendix~\ref{app:euclidean} contains the pole analysis, the conditions
under which the Euclidean denominators stay away from zero, and the
argument for the existence of the continuation.

From here to the end of the section $p$ and $k$ are Euclidean with
$p^{2}\geq0$, the Dirac matrices are Hermitian and satisfy
$\{\gamma_{\mu},\gamma_{\nu}\}=2\delta_{\mu\nu}$, and
$\slashed{k}=\gamma_{\mu}k_{\mu}$. The Euclidean counterpart of the
diagonal Green's functions \eqref{eq:Gstd} is $(i\slashed{k}+M)^{-1}$ with
$M=m$ and $M=m^{*}$, so that the Euclidean Green's function of $\psi$ and
of $\chi$, the counterpart of \eqref{eq:Goriginal}, reads
\begin{equation}\label{eq:SE}
S(k)=\frac{1}{2}\left[\frac{-i\slashed{k}+m}{k^{2}+m^{2}}
+\frac{-i\slashed{k}+m^{*}}{k^{2}+(m^{*})^{2}}\right].
\end{equation}
No overall factor $i$ appears in \eqref{eq:SE}. The conventional factor
$i$ of the $e^{iS}$ normalization belongs to the Minkowski
one-particle-irreducible amplitude and is not carried by the Euclidean
loop coefficient, a point taken up again in the discussion below.

The object we compute is the Euclidean loop coefficient
\begin{equation}\label{eq:Pidef}
\Pi_{E}(p)=a^{2}\int\frac{d^{2\omega}k}{(2\pi)^{2\omega}}\,
\tr\big[S(p+k)\,S(k)\big],
\end{equation}
with $a$ the real Yukawa coupling. The trace taken before integration
keeps the result an $O(4)$ scalar. No termwise Wick rotation and no
Jacobian from a contour deformation enter \eqref{eq:Pidef}.

\subsection{Reality of the traced integrand}

For real Euclidean $k$ and Hermitian $\gamma_{\mu}$,
\begin{equation}\label{eq:conjprop1}
\left(\frac{-i\slashed{k}+M}{k^{2}+M^{2}}\right)^{\dagger}
=\frac{i\slashed{k}+M^{*}}{k^{2}+(M^{*})^{2}},
\end{equation}
which differs from the Green's function of mass $M^{*}$ only in the sign
of the term linear in $\slashed{k}$. Since
$\gamma_{5}\gamma_{\mu}\gamma_{5}=-\gamma_{\mu}$, that sign is restored by
conjugation with $\gamma_{5}$, so we define
\begin{equation}\label{eq:conjdef}
X^{*}\equiv\gamma_{5}\,X^{\dagger}\,\gamma_{5},
\end{equation}
for which
\begin{equation}\label{eq:conjdef2}
\left(\frac{-i\slashed{k}+m}{k^{2}+m^{2}}\right)^{*}
=\frac{-i\slashed{k}+m^{*}}{k^{2}+(m^{*})^{2}},
\end{equation}
together with the same relation with $m$ and $m^{*}$ interchanged. The two
terms of \eqref{eq:SE} are therefore exchanged by \eqref{eq:conjdef}, and
the Euclidean Green's function is self-conjugate,
\begin{equation}\label{eq:Greal2}
\big[S(k)\big]^{*}=S(k).
\end{equation}
The operation \eqref{eq:conjdef} reverses the order of a product,
$(AB)^{*}=B^{*}A^{*}$, and satisfies $\tr(X^{*})=(\tr X)^{*}$, since
$\tr(\gamma_{5}X^{\dagger}\gamma_{5})=\tr(X^{\dagger})$. Hence
\begin{equation*}
\begin{aligned}
\Big\{\tr\big[S(p+k)\,S(k)\big]\Big\}^{*}
&=\tr\big[S(k)^{*}\,S(p+k)^{*}\big]\\
&=\tr\big[S(k)\,S(p+k)\big]\\
&=\tr\big[S(p+k)\,S(k)\big],
\end{aligned}
\end{equation*}
where the last equality uses cyclicity of the Dirac trace. The traced
integrand of \eqref{eq:Pidef} is real at every point of the Euclidean
momentum space and for every real $p$, so that the reality of $\Pi_{E}$ is
a property of the integrand and not of a choice of contour.
Appendix~\ref{app:reality} returns to this point.

\subsection{The integral after the Dirac trace}

\begin{widetext}
The trace identity
$\tr[(-i\slashed{q}_{1}+M_{1})(-i\slashed{q}_{2}+M_{2})]
=4(-q_{1}\cdot q_{2}+M_{1}M_{2})$, applied to the four terms of the
product $S(p+k)S(k)$, turns \eqref{eq:Pidef} into
\begin{equation}\label{eq:PiE}
\begin{aligned}
\Pi_E(p) = a^2 \int \frac{d^{2\omega}k}{(2\pi)^{2\omega}}
\sum_{M_1,M_2}
\frac{-(p+k)\cdot k + M_1 M_2}
{[(p+k)^2 + M_1^2][k^2 + M_2^2]},
\end{aligned}
\end{equation}
where the sum runs over $(M_1,M_2)\in\{(m,m),(m,m^{*}),(m^{*},m),(m^{*},m^{*})\}$ and the factor $\tfrac{1}{4}$ from the two halves of \eqref{eq:SE} cancels against the factor $4$ of the Dirac trace. The denominators of \eqref{eq:PiE} have no zeros for real Euclidean momenta, so dimensional regularization applies directly, as discussed in Appendix~\ref{app:euclidean}.

\subsection{Feynman parametrization}

Feynman parametrization combines each term through
\begin{equation}\label{eq:feynman}
\frac{1}{[(p+k)^2 + M_1^2][k^2 + M_2^2]} = \int_0^1 dx\, \frac{1}{[q^2 + \Delta(p,x,M_1^2,M_2^2)]^2}, \qquad q = k + xp,
\end{equation}
with
\begin{equation}\label{eq:Deltagen}
\Delta(p,x,M_1^2,M_2^2) = x M_1^2 + (1-x) M_2^2 + x(1-x)p^2 .
\end{equation}
The four mass pairs give
\begin{align}
\Delta_{mm}(p,x) &\equiv \Delta(p,x,m^2,m^2) = x(1-x)p^2 + m^2, \label{eq:Dmm}\\
\Delta_{mm^{*}}(p,x) &\equiv \Delta(p,x,m^2,(m^{*})^2) = x(1-x)p^2 + x m^2 + (1-x)(m^{*})^2, \label{eq:Dmms}\\
\Delta_{m^{*}m}(p,x) &\equiv \Delta(p,x,(m^{*})^2,m^2) = x(1-x)p^2 + x(m^{*})^2 + (1-x) m^2, \label{eq:Dmsm}\\
\Delta_{m^{*}m^{*}}(p,x) &\equiv \Delta(p,x,(m^{*})^2,(m^{*})^2) = x(1-x)p^2 + (m^{*})^2. \label{eq:Dmsms}
\end{align}
The numerator rearranges as
\begin{equation}\label{eq:pkk}
(p+k)\cdot k = (q + (1-x)p) \cdot (q - xp) = q^2 - x(1-x)p^2 + (1-2x)p\cdot q,
\end{equation}
and the term linear in $q$ vanishes on integration, so inside the integral
\begin{equation}\label{eq:pkkint}
-(p+k)\cdot k + M_1 M_2 \;\longrightarrow\; -\big(q^2 + \Delta\big) + K(p,x,M_1,M_2),
\end{equation}
where
\begin{equation}\label{eq:Kgen}
K(p,x,M_1,M_2) = \Delta(p,x,M_1^2,M_2^2) + M_1 M_2 + x(1-x)p^2
= x M_1^2 + (1-x)M_2^2 + M_1M_2 + 2x(1-x)p^2 .
\end{equation}
The four pairs give
\begin{align}
K_{mm} &= 2m^2 + 2x(1-x)p^2, \label{eq:Kmm}\\
K_{mm^{*}} &= m m^{*} + x m^2 + (1-x)(m^{*})^2 + 2x(1-x)p^2, \label{eq:Kmms}\\
K_{m^{*}m} &= m m^{*} + x (m^{*})^2 + (1-x) m^2 + 2x(1-x)p^2, \label{eq:Kmsm}\\
K_{m^{*}m^{*}} &= 2(m^{*})^2 + 2x(1-x)p^2. \label{eq:Kmsms}
\end{align}
The integral becomes
\begin{equation}\label{eq:Piq}
\Pi_E(p) = a^2 \int_0^1 dx \int \frac{d^{2\omega}q}{(2\pi)^{2\omega}}
\sum_{M_1,M_2} \frac{-\big(q^2 + \Delta\big) + K}{[q^2 + \Delta]^2}.
\end{equation}
With the abbreviation
\begin{equation}\label{eq:Ldef}
L(\Delta) = \frac{1}{\varepsilon} + \ln\frac{4\pi\Lambda^2}{\Delta} - \gamma ,
\end{equation}
the two Euclidean master integrals read
\begin{equation}\label{eq:masters}
\begin{aligned}
\int \frac{d^{2\omega}q}{(2\pi)^{2\omega}} \frac{1}{[q^2+\Delta]^2}
&= \frac{L(\Delta)}{16\pi^2}, \\[4pt]
\int \frac{d^{2\omega}q}{(2\pi)^{2\omega}} \frac{1}{q^2+\Delta}
&= -\frac{\Delta}{16\pi^2}\big[ L(\Delta) + 1 \big],
\end{aligned}
\end{equation}
and a direct evaluation gives
\begin{equation}\label{eq:dimexp}
\begin{aligned}
\Pi_E(p) = \frac{a^2}{16\pi^2}& \int_0^1 dx \Big[
K_{mm} \Big( \tfrac{1}{\varepsilon} + \ln\tfrac{4\pi \Lambda^2}{\Delta_{mm}} - \gamma \Big) + \Delta_{mm} \Big( \tfrac{1}{\varepsilon} + \ln\tfrac{4\pi \Lambda^2}{\Delta_{mm}} - (\gamma - 1) \Big) \\
+\ &K_{mm^{*}} \Big( \tfrac{1}{\varepsilon} + \ln\tfrac{4\pi \Lambda^2}{\Delta_{mm^{*}}} - \gamma \Big) + \Delta_{mm^{*}} \Big( \tfrac{1}{\varepsilon} + \ln\tfrac{4\pi \Lambda^2}{\Delta_{mm^{*}}} - (\gamma - 1) \Big) \\
+\ &K_{m^{*}m} \Big( \tfrac{1}{\varepsilon} + \ln\tfrac{4\pi \Lambda^2}{\Delta_{m^{*}m}} - \gamma \Big) + \Delta_{m^{*}m} \Big( \tfrac{1}{\varepsilon} + \ln\tfrac{4\pi \Lambda^2}{\Delta_{m^{*}m}} - (\gamma - 1) \Big) \\
+\ &K_{m^{*}m^{*}} \Big( \tfrac{1}{\varepsilon} + \ln\tfrac{4\pi \Lambda^2}{\Delta_{m^{*}m^{*}}} - \gamma \Big) + \Delta_{m^{*}m^{*}} \Big( \tfrac{1}{\varepsilon} + \ln\tfrac{4\pi \Lambda^2}{\Delta_{m^{*}m^{*}}} - (\gamma - 1) \Big) \Big] + \mathcal{O}(\varepsilon).
\end{aligned}
\end{equation}
The pole part reads
\begin{equation}\label{eq:Pidiv}
(\Pi_E)_{\text{div}}(p) = \frac{a^2}{16\pi^2 \varepsilon} \int_0^1 dx \Big[
K_{mm} + \Delta_{mm} + K_{mm^{*}} + \Delta_{mm^{*}}
+ K_{m^{*}m} + \Delta_{m^{*}m} + K_{m^{*}m^{*}} + \Delta_{m^{*}m^{*}} \Big],
\end{equation}
and is absorbed into a redefinition of the scalar fields in the momentum and mass sectors and of the scalar mass term. The finite part reads
\begin{equation}\label{eq:Pifin}
\begin{aligned}
(\Pi_E)_{\text{fin}}(p) = \frac{a^2}{16\pi^2} \int_0^1 dx \Big[
&K_{mm} \Big( \ln\tfrac{4\pi \Lambda^2}{\Delta_{mm}} - \gamma \Big) + \Delta_{mm} \Big( \ln\tfrac{4\pi \Lambda^2}{\Delta_{mm}} - (\gamma - 1) \Big) \\
+\ &K_{mm^{*}} \Big( \ln\tfrac{4\pi \Lambda^2}{\Delta_{mm^{*}}} - \gamma \Big) + \Delta_{mm^{*}} \Big( \ln\tfrac{4\pi \Lambda^2}{\Delta_{mm^{*}}} - (\gamma - 1) \Big) \\
+\ &K_{m^{*}m} \Big( \ln\tfrac{4\pi \Lambda^2}{\Delta_{m^{*}m}} - \gamma \Big) + \Delta_{m^{*}m} \Big( \ln\tfrac{4\pi \Lambda^2}{\Delta_{m^{*}m}} - (\gamma - 1) \Big) \\
+\ &K_{m^{*}m^{*}} \Big( \ln\tfrac{4\pi \Lambda^2}{\Delta_{m^{*}m^{*}}} - \gamma \Big) + \Delta_{m^{*}m^{*}} \Big( \ln\tfrac{4\pi \Lambda^2}{\Delta_{m^{*}m^{*}}} - (\gamma - 1) \Big) \Big].
\end{aligned}
\end{equation}
\end{widetext}

\subsection{Discussion}

The mass pairs satisfy
\begin{equation}\label{eq:conjpairs}
\Delta_{m^{*}m^{*}} = (\Delta_{mm})^{*}, \qquad \Delta_{m^{*}m} = (\Delta_{mm^{*}})^{*},
\end{equation}
and the same pairing holds for the numerator coefficients,
\begin{equation}\label{eq:conjpairsK}
K_{m^{*}m^{*}} = (K_{mm})^{*}, \qquad K_{m^{*}m} = (K_{mm^{*}})^{*},
\end{equation}
since $m m^{*} = |m|^2$ and $p^{2}$ are real. The four terms in \eqref{eq:dimexp} group into two complex conjugate pairs, so the $x$-integral equals twice the real part of the $(m,m)$ and $(m,m^{*})$ terms. Thus $\Pi_E$ is real throughout the Euclidean domain, and its analytic continuation remains real on the real axis below the first mixed-pair threshold. The pairing already holds at the level of the integrand of \eqref{eq:PiE}, at every point of the Euclidean momentum space, so that the reality of the result is a property of the integrand and not of a choice of contour, as developed in Appendix~\ref{app:euclidean}.

In the Minkowski normalization, the factors from the two Yukawa vertices, two fermion propagators, and the closed fermion loop give
\begin{equation}\label{eq:icount}
(ia)(ib)\,(i)(i)\,(-1)=-ab\;\overset{b=-a}{=}\;a^{2}.
\end{equation}
No additional factor $i$ is assigned to the Euclidean integration measure. The analytic continuation of $\Pi_E$ defines the scalar loop function $\Sigma(s)$, while the Minkowski one-particle-irreducible amplitude carries the conventional overall factor $i$ associated with the $e^{iS}$ normalization. For real $s<4N^{2}$, $\Sigma(s)$ is real. Above threshold, its boundary values satisfy $\Sigma(s-i0)=\Sigma(s+i0)^{*}$, and the spectral density
\[
\rho(s)=\frac{1}{2\pi i}\,\mathrm{Disc}\,\Sigma(s)
\]
is therefore real. This statement concerns the two-propagator loop considered here. Loops with three or more internal lines require a separate analysis of the analytic structure of the complex-conjugate poles.

\textit{Analytic structure and the location of the thresholds.} It is instructive to contrast the present result with the familiar case of physical fermions of real mass $m_{\rm phys}$. There, the scalar two-point function acquires an imaginary part for $p^{2}>4m_{\rm phys}^{2}$, corresponding to the decay $\phi\to\bar\psi\psi$ into asymptotic states and enforced by the optical theorem. In our model the fermions are not asymptotic states: their propagators carry the complex conjugate poles $m_{\pm}=N\pm iav$ and are the analogues of the $i$-particles of the Gribov--Zwanziger framework. The thresholds separate accordingly. The two equal-mass pairs place theirs at $p^{2}=4m^{2}$ and $p^{2}=4(m^{*})^{2}$, both away from the real axis, and contribute no discontinuity for real $p$; this is the precise sense in which the naive two-fermion threshold has been displaced into the complex plane. The mixed pairs, on the other hand, are built out of constituents whose masses add up to the real number $m+m^{*}=2N$, and place their threshold at
\begin{equation}\label{eq:threshold}
p^{2}=(m+m^{*})^{2}=4N^{2},
\end{equation}
on the real axis, where the Feynman parameter contour is pinched, as shown in Appendix~\ref{app:euclidean}. A cut therefore opens at $4N^{2}$. The two boundary values of $\Sigma$ across this cut are complex conjugates, so $\mathrm{Disc}\,\Sigma(s)=2i\,\mathrm{Im}\,\Sigma(s+i0)$ and the associated spectral density $\rho(s)=(2\pi i)^{-1}\mathrm{Disc}\,\Sigma(s)$ is real. Positivity of $\rho$ is not established by the present calculation. The cut at $4N^{2}$ is not the Cutkosky cut of a decay $\phi\to\bar\psi\psi$ into asymptotic fermion states, because the constituent fermions have complex poles. It is the real spectral cut of the two-propagator loop generated by the mixed complex-conjugate pair, in the same analytic sense in which correlators built from $i$-particle pairs admit real spectral representations \cite{Dudal:2010wn,iparticles}.

\section{Gauge sector}

The gauge and scalar part of the action reads
\begin{multline}\label{eq:gaugeaction}
S = \int d^4x\, \Big[ -\tfrac{1}{4} f_{\mu\nu}(a)f^{\mu\nu}(a) -\tfrac{1}{4} f_{\mu\nu}(\hat{a})f^{\mu\nu}(\hat{a}) \\
+(D_\mu \phi)^\dagger (D^\mu \phi) - V(\phi^\dagger \phi) \Big],
\end{multline}
and a scalar expectation value gives mass terms for the gauge fields,
\begin{multline}\label{eq:massgauge}
(D_\mu \phi)^\dagger (D^\mu \phi) \rightarrow (D_\mu \phi)^\dagger (D^\mu \phi) + iev A_{\mu}\big((D^{\mu}\phi)-(D^\mu \phi)^{\dagger}\big) \\ + e^{2}v^{2}A_{\mu}A^{\mu},
\end{multline}
where $A_{\mu} = a_{\mu} + \hat{a}_{\mu}$. The mass falls on $A_{\mu}=a_{\mu}+\hat{a}_{\mu}$, so $B_{\mu}=a_{\mu}-\hat{a}_{\mu}$ stays massless. A consistent calculation needs gauge fixing. A 't Hooft-type gauge removes the mixed bilinear between $A_{\mu}$ and the scalars, and a Landau-type gauge handles $B_{\mu}$. BRST symmetry builds the gauge fixing through a pair of doublets,
\begin{equation}\label{eq:doublets}
s\overline{c}=b,\quad sb=0, \qquad s\hat{\overline{c}}=\hat{b},\quad s\hat{b}=0,
\end{equation}
and the gauge-fixing functional
\begin{equation}\label{eq:gaugefix}
\begin{aligned}
S_{gf}=bG&+\frac{\alpha}{2}b^{2}+ \hat{b}\hat{G} + \hat{\overline{c}}\,\partial^{2}(c+\hat{c})\\ &+ \overline{c}(\partial^{2}+ 2e^{2}v^{2}\alpha)(c-\hat{c})\\ &+ e^{2}v\alpha (c-\hat{c})(\phi+\phi^{\dagger}) \\
\end{aligned}
\end{equation}
where
\begin{equation}
G = \partial^{\mu}A_{\mu}+iev\alpha(\phi-\phi^{\dagger}), \quad \hat{G}=\partial^{\mu}B_{\mu}.
\end{equation}
Integration over $b$ turns $bG+\frac{\alpha}{2}b^{2}$ into $-\frac{1}{2\alpha}G^{2}$ and removes the mixed term between the scalars and the sum of the gauge fields. The BRST transformations before and after symmetry breaking differ. Before breaking they read
\begin{equation}\label{eq:brstbefore}
\begin{aligned}
s a_\mu &= -\partial_\mu c, & s \hat{a}_\mu &= \partial_\mu \hat{c}, \\
s c &= 0, & s \hat{c} &= 0, \\
s \psi &= -ie\,c\,\psi, & s \overline{\psi} &= ie\,\overline{\psi}\,c, \\
s \chi &= -ie\,\hat{c}\,\chi, & s \overline{\chi} &= ie\,\overline{\chi}\,\hat{c}, \\
s \phi &= ie\,(\hat{c} - c)\,\phi, & s \phi^\dagger &= -ie\,(\hat{c} - c)\,\phi^\dagger,
\end{aligned}
\end{equation}
and after breaking
\begin{equation}\label{eq:brstafter}
\begin{aligned}
&s a_\mu = -\partial_\mu c, & s \hat{a}_\mu = \partial_\mu \hat{c}, \\
&s c = 0, & s \hat{c} = 0, \\
&s \psi = -ie\,c\,\psi, & s \overline{\psi} = ie\,\overline{\psi}\,c, \\
&s \chi = -ie\,\hat{c}\,\chi, & s \overline{\chi} = ie\,\overline{\chi}\,\hat{c}, \\
&s \phi = ie\,(\hat{c} - c)\,\phi +ie\,(\hat{c} - c)\,v, \\
&s \phi^\dagger = -ie\,(\hat{c} - c)\,\phi^\dagger -ie\,(\hat{c} - c)\,v,
\end{aligned}
\end{equation}
which stays nilpotent, since $c^{2}=0$, $\hat{c}^{2}=0$, and $\{c,\hat{c}\}=0$, and leaves the broken-phase action invariant. The broken symmetry is the one of the chosen vacuum, and feeding this into the transformation gives a nilpotent symmetry of the action with the new vacuum. The change in the cohomology follows.

\subsection{BRST symmetry in the broken vacuum}

A filtration expands $s$ as $s=s_{0}+e\,s_{1}$. Nilpotency imposes
\begin{equation*}
s = s_{0}+e\,s_{1}, \quad s^{2}=0 \ \Rightarrow\ (s_{0})^{2}=0, \ \{s_{0},s_{1}\}=0, \ (s_{1})^{2}=0,
\end{equation*}
and $s_{0}\phi=0$ before breaking, with the same for the conjugate. After breaking
\begin{equation}\label{eq:s0phi}
s_{0}\phi = ie\,(\hat{c} - c)\,v, \qquad s_{0}\phi^{\dagger}= -ie\,(\hat{c} - c)\,v.
\end{equation}
With $A_{\mu}=-\partial_{\mu}(c-\hat{c})$ under $s_0$, two combinations are $s_{0}$-invariant,
\begin{equation}\label{eq:invcomb}
s_{0}(\partial_{\mu}\phi -ievA_{\mu})=0, \qquad s_{0}(\partial_{\mu}\phi^{\dagger}+ievA_{\mu})=0,
\end{equation}
and behave as matter fields with a mass. The cohomology of $s$ is isomorphic to a subspace of the cohomology of $s_{0}$, so a combination of $A_{\mu}$ with the scalars gains mass as matter, while $B_{\mu}$ stays massless.

\begin{widetext}
\subsubsection{Gauge Green's functions}

The quadratic gauge action reads
\begin{equation}\label{eq:gaugeS}
S = \int d^4x \left[ -\tfrac18 F_{\mu\nu}(A)F^{\mu\nu}(A) -\tfrac18 F_{\mu\nu}(B)F^{\mu\nu}(B) - \frac{1}{2\alpha}(\partial_\mu A^\mu)^2 + e^2v^2 A_\mu A^\mu + \hat{b}\,\partial^{\mu}B_{\mu} \right],
\end{equation}
with $F_{\mu\nu}(A) = \partial_\mu A_\nu - \partial_\nu A_\mu$ and the same for $B$. The propagators are
\begin{align}
G(A^\mu(p) A^\nu(-p)) &= \frac{-2\theta^{\mu\nu}}{p^2 - 4e^2v^2} - \frac{\alpha}{p^2 - 2\alpha e^2v^2}\, \omega^{\mu\nu}, \label{eq:GAA}\\
G(B^\mu(p) B^\nu(-p)) &= -\frac{2}{p^2}\, \theta^{\mu\nu}(p), \label{eq:GBB}
\end{align}
with the projectors
\begin{equation}\label{eq:proj}
\theta^{\mu\nu} = \eta^{\mu\nu} - \frac{p^\mu p^\nu}{p^2}, \qquad \omega^{\mu\nu} = \frac{p^\mu p^\nu}{p^2},
\end{equation}
and, for the original fields,
\begin{align}
G(a^\mu(p) a^\nu(-p)) &= -\frac{\theta^{\mu\nu}}{2(p^2 - 4e^2v^2)} - \frac{\theta^{\mu\nu}}{2p^2} - \frac{\alpha}{4}\frac{\omega^{\mu\nu}}{p^2 - 2\alpha e^2v^2}, \label{eq:Gaa}\\
G(\hat{a}^\mu(p) \hat{a}^\nu(-p)) &= G(a^\mu a^\nu), \label{eq:Gaah}\\
G(a^\mu(p) \hat{a}^\nu(-p)) &= \frac{\theta^{\mu\nu}}{2p^2} - \frac{\theta^{\mu\nu}}{2(p^2 - 4e^2v^2)} - \frac{\alpha}{4}\frac{\omega^{\mu\nu}}{p^2 - 2\alpha e^2v^2}. \label{eq:Gahat}
\end{align}
A term $F_{\mu\nu}(a)F^{\mu\nu}(\hat{a})$ cannot be excluded. A non-Abelian theory suppresses it, since the curvature there transforms covariantly, while the Abelian curvature is invariant and the term survives. In terms of $A_{\mu}$ and $B_{\mu}$ the general action reads
\begin{equation}\label{eq:xaction}
\begin{aligned}
S = \int d^4x \Big[ &\frac{\zeta-1}{8} F_{\mu\nu}(A)F^{\mu\nu}(A) - \frac{\zeta+1}{8} F_{\mu\nu}(B)F^{\mu\nu}(B) \\
&- \frac{1}{2\alpha}(\partial_\mu A^\mu)^2 + e^2v^2 A_\mu A^\mu + \hat{b}\,\partial^\mu B_\mu \Big],
\end{aligned}
\end{equation}
where $\zeta$ carries the mixed term $F_{\mu\nu}(a)F^{\mu\nu}(\hat{a})$, a non-trivial BRST cocycle. Its presence enlarges the local BRST-invariant basis at ultraviolet dimension $\leq4$; a complete classification of the allowed counterterms requires the full algebraic renormalization analysis \cite{Piguet:1995er}. The propagators are
\begin{align}
G(A^\mu(p) A^\nu(-p)) &= -\frac{2}{1-\zeta}\frac{\theta^{\mu\nu}}{p^2 - \frac{4e^2v^2}{1-\zeta}} - \frac{\alpha}{p^2 - 2\alpha e^2v^2}\,\omega^{\mu\nu}, \label{eq:GAAx}\\
G(B^\mu(p) B^\nu(-p)) &= -\frac{2}{(1+\zeta)p^2}\,\theta^{\mu\nu}, \label{eq:GBBx}
\end{align}
with the projectors of \eqref{eq:proj}. The combinations
\begin{equation}\label{eq:aacomb}
a_{\mu} = \tfrac{1}{2}(A_{\mu}+B_{\mu}), \qquad \hat{a}_{\mu} = \tfrac{1}{2}(A_{\mu}-B_{\mu}),
\end{equation}
give the original-field propagators
\begin{align}
G(a^\mu(p) a^\nu(-p)) &= -\frac{\theta^{\mu\nu}}{2(1-\zeta)\left(p^2 - \frac{4e^2v^2}{1-\zeta}\right)} - \frac{\theta^{\mu\nu}}{2(1+\zeta)p^2} - \frac{\alpha}{4}\frac{\omega^{\mu\nu}}{p^2 - 2\alpha e^2v^2}, \label{eq:Gaax}\\
G(\hat{a}^\mu(p) \hat{a}^\nu(-p)) &= G(a^\mu(p) a^\nu(-p)), \label{eq:Gaahx}\\
G(a^\mu(p) \hat{a}^\nu(-p)) &= -\frac{\theta^{\mu\nu}}{2(1-\zeta)\left(p^2 - \frac{4e^2v^2}{1-\zeta}\right)} + \frac{\theta^{\mu\nu}}{2(1+\zeta)p^2} - \frac{\alpha}{4}\frac{\omega^{\mu\nu}}{p^2 - 2\alpha e^2v^2}. \label{eq:Gahatx}
\end{align}
\end{widetext}

Massive propagation needs $\zeta<1$ and massless propagation needs $\zeta>-1$. The action is admissible for $-1<\zeta<1$, while the residue of the mixed propagator $\langle a_{\mu}\hat{a}_{\nu}\rangle$ marks a non-physical mode, a ghost. The gauge used here has a ghost sector coupled to the scalars and gauge fields, as required by the BRST gauge-fixing construction. No independent unitarity statement is inferred from this coupling. The bilinear ghost sector
\begin{equation}\label{eq:ghostbil}
\mathcal{L}_{\text{ghost}}^{\text{bilinear}} = \bar{c}(\partial^{2} + 2\alpha e^{2}v^{2})(c - \hat{c}) + \hat{\bar{c}}\,\partial^{2}(c + \hat{c}),
\end{equation}
gives the propagators
\begin{equation}
\begin{aligned}
G(\bar{c}(p)\, c(-p)) &= -\frac{1}{2(p^{2} - 2\alpha e^{2}v^{2})}, \\
G(\hat{\bar{c}}(p)\, c(-p)) & = -\frac{1}{2p^{2}}, \label{eq:ghostG1}\\
\end{aligned}
\end{equation}
\vspace{-1em}
\begin{equation}
\begin{aligned}
G(\bar{c}(p)\, \hat{c}(-p)) & = \frac{1}{2(p^{2} - 2\alpha e^{2}v^{2})}, \\
G(\hat{\bar{c}}(p)\, \hat{c}(-p)) & = -\frac{1}{2p^{2}}. \label{eq:ghostG2}
\end{aligned}
\end{equation}
The action displayed above is BRST invariant and power-counting renormalizable. The determination of the counterterm basis and the stability analysis through algebraic renormalization are separate work. The fermion loop with complex-conjugate poles gives the real scalar two-point coefficient described above.

\section{Conclusion}

The model assembled here couples two fermions, two Abelian gauge fields, and one complex scalar within a single BRST symmetry, and the tuning $a=-b$ drives the fermion mass matrix outside the Hermitian class, so that the propagators acquire the complex conjugate poles $m=N\pm iav$. Against this background the main computation shows that the one-loop fermion contribution to the scalar two-point function $\langle\phi^{\dagger}(p)\phi(-p)\rangle$ stays real for real external momentum below the threshold $(m+m^{*})^{2}=4N^{2}$, once the conventional overall factor $i$ of the Minkowski one-particle-irreducible amplitude is separated from the Euclidean loop coefficient. The thresholds of the equal-mass pairs are carried into the complex $p^{2}$ plane and leave the real axis untouched, while the mixed pairs, whose constituent masses add up to the real number $2N$, open a cut at $4N^{2}$. Across this cut the boundary values are related by complex conjugation and define a real spectral density, whose positivity is not established here. The present calculation is restricted to the two-propagator loop. Higher composite powers require a separate spectral analysis.

Two features give the construction its reach. The non-Hermiticity is confined to the elementary fermion propagators, whereas the mixed two-propagator channel has a real Euclidean loop coefficient below threshold and a real spectral density after continuation across its real cut \cite{Dudal:2010wn}. This realizes within a Yukawa and Higgs setting the pattern already identified for the $i$-particles of the Gribov-Zwanziger gluon \cite{iparticles,Capri2011, Capri:2011ki} and for the indefinite-metric excitations of Lee and Wick \cite{LeeWick1969}, the same complex-pole structure that underlies the non-Hermitian platforms now explored in condensed matter \cite{Ge2024,Xu2024}. The present action is BRST invariant and power-counting renormalizable, while the determination of the counterterm basis still requires the algebraic renormalization analysis \cite{Piguet:1995er}.

A non-Hermitian theory does not secure unitarity on its own. What the reality of the $\phi^{\dagger}\phi$ correlator establishes is that composite observables can stay real even when the elementary excitations carry complex conjugate masses, the behavior one would expect on a subspace where the evolution is unitary with respect to a suitable metric, although the present one-loop result does not by itself establish that such a metric exists. The explicit construction of that metric, the analysis of the composite operator $\overline{\Psi}_{+}\Psi_{-}$ beyond one loop, and the completion of the algebraic renormalization program mark the natural continuation of this work.

\appendix

\section{The loop integral in Euclidean space and its continuation to real momenta}
\label{app:euclidean}

The fermion propagators of the broken phase carry the complex conjugate
masses $m=N+iav$ and $m^{*}=N-iav$, and the analytic structure that
follows from them is not the one of an ordinary Dirac field. We devote
this appendix to the definition of the loop integral of Sec.~IV, since the
four terms of \eqref{eq:PiE} have no common Minkowski counterpart that a
single rotation of the integration contour would reach, and the reality of
the result does not rest on such a rotation.

\subsection{Why a global rotation of the Minkowski contour is not available}
\label{app:nowick}

We begin with the position of the poles. For a denominator $k^{2}-M^{2}$
with $M^{2}=\mu^{2}+i\delta$ and $\delta\neq0$, the singularities in the
complex $k^{0}$ plane sit at
\begin{equation}\label{eq:polepos}
k^{0}=\pm\,\omega,\qquad \omega=\sqrt{\vec{k}^{\,2}+M^{2}},
\end{equation}
the square root being taken on its principal branch, which preserves the
sign of the imaginary part of its argument. Then
$\mathrm{sgn}(\mathrm{Im}\,\omega)=\mathrm{sgn}(\delta)$ for every
$\vec{k}$, and, with
\begin{equation}\label{eq:masssquares}
m^{2}=N^{2}-a^{2}v^{2}+2iNav,\qquad
(m^{*})^{2}=N^{2}-a^{2}v^{2}-2iNav,
\end{equation}
for $Nav>0$ (the opposite sign only interchanges the two conjugate assignments), the two poles of the $m^{2}$ propagator lie in the first and third
quadrants of the complex $k^{0}$ plane and the two poles of the
$(m^{*})^{2}$ propagator in the second and fourth. Only the second
configuration is the one familiar from a real mass supplemented by the
Feynman prescription, and only for it does the counterclockwise contour
$k^{0}\to e^{i\theta}k^{0}_{E}$, $\theta:0\to\pi/2$, sweep a region free
of singularities. In fact, a rotation performed in the same sense on the
$m^{2}$ propagator crosses its poles. The obstruction is sharpest in the
Minkowski integrand that would correspond to \eqref{eq:PiE},
\begin{equation}\label{eq:PiM}
\sum_{M_{1},M_{2}}
\frac{(p+k)\cdot k+M_{1}M_{2}}
{[(p+k)^{2}-M_{1}^{2}][k^{2}-M_{2}^{2}]},
\end{equation}
whose two mixed terms carry the four poles together and admit no sense of
rotation at all. One should therefore not read \eqref{eq:PiE} as the
result of a single Wick rotation applied to a Minkowski integral, and we
do not define it in that way.

\subsection{The loop integral is defined in Euclidean space}
\label{app:euclid}

The definition we adopt is the one already in use for composite operators
built out of constituents with complex conjugate poles
\cite{iparticles,Dudal:2010wn}: the loop integral is the Euclidean one,
\eqref{eq:PiE}, and the dependence on real external momentum is obtained
from it by analytic continuation. The virtue of this definition is that no
contour is deformed and no pole is approached. The domain of integration is
the real Euclidean momentum space, and the denominators never vanish on it.
In fact, from \eqref{eq:Dmm}--\eqref{eq:Dmsms},
\begin{equation}\label{eq:ImDelta}
\mathrm{Im}\,\Delta_{mm}=2Nav,\qquad
\mathrm{Im}\,\Delta_{mm^{*}}=2Nav\,(2x-1),
\end{equation}
together with the conjugate relations \eqref{eq:conjpairs}, so that the
imaginary part of $\Delta$ is different from zero for every $x\in[0,1]$ in
the two pure terms, and vanishes in the two mixed terms only at $x=1/2$,
where
\begin{equation}\label{eq:Deltahalf}
\Delta_{mm^{*}}\Big|_{x=1/2}=\Delta_{m^{*}m}\Big|_{x=1/2}
=N^{2}-a^{2}v^{2}+\frac{p^{2}}{4},
\end{equation}
a quantity that is strictly positive in the regime \eqref{eq:regime}.
Under \eqref{eq:regime} the four denominators of \eqref{eq:Piq} are
regular on the whole Euclidean domain, the ultraviolet behaviour is the
usual one, and dimensional regularization applies without further
qualification.

\subsection{Reality of the integrand}
\label{app:reality}

The reality of the fermion contribution is a property of the Euclidean traced integrand and not of the way a contour is chosen. Under the operation defined in \eqref{eq:conjdef}, the Euclidean Green's function is self-conjugate, $S^{*}=S$. Because this conjugation is an anti-automorphism, $(AB)^{*}=B^{*}A^{*}$, cyclicity of the Dirac trace restores the original order in the traced product. Equivalently, the four terms of \eqref{eq:PiE} are exchanged in pairs by complex conjugation: the $(m,m)$ term is the conjugate of the $(m^{*},m^{*})$ term, and the $(m,m^{*})$ term is the conjugate of the $(m^{*},m)$ term. This holds at every point of Euclidean momentum space and for every Feynman parameter, since $mm^{*}=|m|^{2}$ is real. The integrand of \eqref{eq:dimexp} therefore equals twice the real part of its first and second lines, and $\Pi_E$ is real on the Euclidean domain. No contour rotation enters this argument.

\subsection{Continuation to real external momentum}
\label{app:continuation}

It remains to follow the continuation $p_{E}^{2}\to-p^{2}$ of the
functions \eqref{eq:Pifin} and to locate its singularities. A singularity
appears where a zero of $\Delta(x)$ traps the Feynman parameter contour,
so we track the zeros as $p^{2}$ runs along the real axis.

The two pure terms may be disposed of at once. Their imaginary parts are
the constants $\pm2Nav$ of \eqref{eq:ImDelta}, so $\Delta_{mm}$ and
$\Delta_{m^{*}m^{*}}$ never vanish for real $p^{2}$ and for real $x$; the
coincidence of their two zeros occurs at $p^{2}=4m^{2}$ and
$p^{2}=4(m^{*})^{2}$, away from the real axis. The equal-mass thresholds
are in this sense displaced into the complex plane, and the pure terms
remain analytic, and their sum real, for every real $p^{2}$.

The mixed terms behave differently, and it is worth following them in
detail. Writing
\begin{equation}\label{eq:Deltaquad}
\Delta_{mm^{*}}(x)=p^{2}x^{2}+\big(m^{2}-(m^{*})^{2}-p^{2}\big)x+(m^{*})^{2},
\end{equation}
the sum and the product of its two zeros give, for real $p^{2}$, two roots
$x_{\pm}=\tfrac{1}{2}+iy_{\pm}$ placed symmetrically about the middle of
the interval, with
\begin{equation}\label{eq:yroots}
y_{+}+y_{-}=-\frac{4Nav}{p^{2}},\qquad
y_{+}y_{-}=\frac{1}{4}-\frac{N^{2}-a^{2}v^{2}}{p^{2}}.
\end{equation}
At $p^{2}=4(N^{2}-a^{2}v^{2})$ the product vanishes: one of the roots
reaches the real axis at $x=1/2$, and the contour must be deformed
downwards to let it pass. From there on both roots lie below the real
axis, at heights $y_{\pm}<0$, and the deformed contour runs between them.
They approach each other as $p^{2}$ grows and meet at
\begin{equation}\label{eq:pinch}
p^{2}=(m+m^{*})^{2}=4N^{2},\qquad
x_{*}=\frac{m^{*}}{m+m^{*}}=\frac{1}{2}-\frac{iav}{2N},
\end{equation}
where the contour, caught between them, is pinched. This is a genuine
singularity of the continuation, and it sits on the real $p^{2}$ axis:
although each constituent mass is complex, the two masses of a mixed pair
add up to the real number $m+m^{*}=2N$. The remaining coincidence of the
zeros, the pseudo-threshold $p^{2}=(m-m^{*})^{2}=-4a^{2}v^{2}$, occurs at
$x_{*}=\tfrac{1}{2}+iN/(2av)$, above the real axis, where the contour is
never taken, and produces no singularity.

Two consequences follow. The continuation of \eqref{eq:Pifin} is analytic on the real axis for $p^{2}<4N^{2}$, with the Euclidean region included, and the continued scalar coefficient $\Sigma(p^{2})$ is real there by the pairing of Appendix~\ref{app:reality}. At $p^{2}=4N^{2}$ a cut opens. The reflection property gives $\Sigma(s-i0)=\Sigma(s+i0)^{*}$ for real $s$ on the cut, so
\[
\mathrm{Disc}\,\Sigma(s)=2i\,\mathrm{Im}\,\Sigma(s+i0),
\qquad
\rho(s)=\frac{1}{2\pi i}\mathrm{Disc}\,\Sigma(s)\in\mathbb{R}.
\]
This is the spectral structure relevant to the complex-conjugate mass pair discussed in~\cite{Dudal:2010wn}.

One should notice, finally, that the value $p^{2}=4(N^{2}-a^{2}v^{2})$ at
which the first root crosses the contour is \emph{not} a singularity of
$\Sigma$. It is the value at which the curve traced by $\Delta_{mm^{*}}(x)$,
as $x$ runs over the interval, begins to cross the negative real axis, so
that the principal branch of the logarithm in \eqref{eq:Pifin} ceases to
be the continuous one and, if kept, introduces a spurious jump of $i\pi$.
With the contour deformed as above, and the branch followed continuously,
$\Sigma$ passes through that value smoothly.

Equation~\eqref{eq:PiE} is defined directly in Euclidean space and therefore contains no Jacobian from a contour rotation. The factors from the Yukawa vertices, fermion propagators, and the closed fermion loop are displayed in \eqref{eq:icount}; no factor $i$ is assigned to the Euclidean measure. The analytic continuation of $\Pi_E$ defines the scalar coefficient $\Sigma(s)$. The corresponding Minkowski one-particle-irreducible amplitude carries the conventional overall factor $i$ in the $e^{iS}$ normalization. The coefficient $\Sigma(s)$ is real for real $s<4N^{2}$, while above threshold its two boundary values are complex conjugates and determine the real spectral density written above.

\end{document}